\documentclass[aps,prd,amsmath,amssymb,eqsecnum,nofootinbib]{revtex4}
\usepackage{graphicx}

\begin{document}
\title{Extraction of bound-state parameters from dispersive sum rules}
\author{Wolfgang Lucha}%W.~Lucha}
\affiliation{Institute for High Energy Physics, Austrian Academy of Sciences, Austria}
\author{Dmitri Melikhov}%D.~Melikhov}
\affiliation{SINP, Moscow State University, Russia}
\affiliation{Institute for High Energy Physics, Austrian Academy of Sciences, Austria}
\affiliation{Faculty of Physics, University of Vienna, Austria}
\author{Silvano Simula}%S.~Simula}
\affiliation{INFN, Sezione di Roma III, Italy}
\begin{abstract}
The procedure of extracting the ground-state parameters from
vacuum-to-vacuum and vacuum-to-hadron correlators within the
method of sum rules is considered. The emphasis is laid on the
crucial ingredient of this method --- the effective continuum
threshold. A new algorithm to fix this quantity is proposed and
tested. First, a quantum-mechanical potential model which provides
the only possibility to probe the reliability and the actual
accuracy of the sum-rule method is used as a study case. In this
model, our algorithm is shown to lead to a remarkable improvement
of the accuracy of the extracted ground-state parameters compared
to the standard procedures adopted in the method and used in all
previous applications of dispersive sum rules in QCD. As a next
step, it is demonstrated that the procedures of extracting the
ground-state decay constant in the potential model and in QCD are
quantitatively very close to each other. Therefore, the
application of the proposed algorithm in QCD promises a
considerable increase of the accuracy of the extracted hadron
parameters.
\end{abstract}
\maketitle

\section{Introduction}
The method of dispersive sum rules \cite{svz,ioffe} is one of the widely used methods for
obtaining properties of the ground-state hadrons in QCD. The method involves two steps:
(i) one calculates the relevant correlator in QCD at relatively small values of the Euclidean time;
(ii) one applies numerical procedures prompted by quark-hadron duality in order to extract the ground-state
contribution from this correlator. These numerical procedures cannot determine a single value of the
ground-state parameter but should provide the band of values containing the true
hadron parameter with a flat probability distribution within this band \cite{svz}.
The width of this band is a systematic, or intrinsic, uncertainty of the method of sum rules.

It is often claimed that the standard procedures of the method of
sum rules allow one to obtain hadron parameters with about 10\%
accuracy for normal and exotic hadrons. In many cases this is
true. However, there are some examples where the difference
between the sum-rule predictions and the experimental results or
between the results from different versions of sum rules turn out
to be {\it quite unexpectedly} much larger than 10\%: as the first
example, recall an evident conflict between the prediction for the
$D^*D\pi$ coupling from light-cone sum rules \cite{belyaev} and
the subsequent measurement of CLEO \cite{cleo}; as the second
example, mention the incompatibility of the predictions for the
pion form factor at intermediate momentum transfers from
light-cone sum rules \cite{braun}, local-duality sum rules
\cite{blm_ld}, and sum rules with non-local condensates
\cite{stefanis}. A posteriori, these issues may be clarified. The
real problem is, however, that the large errors
--- well exceeding a comfortable 10\% level ---
could not be foreseen on the basis of the standard criteria adopted in the method of sum rules in QCD.

An unbiased judgement of the reliability of the extraction
procedures adopted in the method of sum rules may be acquired by
applying these procedures to problems where the ground-state parameters
may be found independently and exactly as soon as the parameters
of the theory are fixed. Presently, only quantum-mechanical potential models
provide such a possibility.

A simple harmonic-oscillator (HO) potential model \cite{nsvz1}
possesses two essential features of QCD ---
confinement and asymptotic freedom \cite{nsvz} --- and has the following valuable features:
(i) the bound-state parameters (masses, wave functions, form factors) are known precisely;
(ii) direct analogues of the QCD correlators may be calculated exactly.
(For a discussion of many aspects of sum rules in quantum mechanics we refer to
\cite{qmsr,orsay,ms_inclusive,radyushkin2001}).

Making use of this model, we have already studied the extraction
of ground-state parameters from different types of correlators:
namely, the ground-state decay constant from two-point
vacuum-to-vacuum correlator \cite{lms_2ptsr}, the form factor from
three-point vacuum-to-vacuum correlator \cite{lms_3ptsr}, and the
form factor from vacuum-to-hadron correlator \cite{m_lcsr}. We
have demonstrated that the standard procedures adopted in the
method of sum rules do not work properly: for all types of
correlators the true value of the bound-state parameter was shown
to lie outside the band obtained according to the standard
criteria. These results gave us a solid ground to claim that also
in QCD the actual accuracy of the method turns out to be worse
than expected on the basis of applying the standard criteria.

We have realized that the main origin of this difficulty of the
method lies in an over-simplified model for the hadron continuum,
which is described as a perturbative contribution above a constant
Borel-parameter-independent effective continuum threshold. We have
introduced the notion of the {\it exact} effective continuum
threshold, which corresponds to the true bound-state parameters:
in the HO model the true hadron parameters --- decay constant and
form factor --- are known and the exact effective continuum
thresholds for different correlators may be calculated. We have
demonstrated that the exact effective continuum threshold (i) is
not a universal quantity and depends on the correlator considered
(i.e., it is in general different for two-point and three-point correlators), (ii) depends on the Borel parameter
and, for the form-factor case, also on the momentum transfer.

Moreover, we have shown that the ``Borel stability criterion''
combined with the assumption of a Borel-parameter-independent
effective continuum threshold leads to the extraction of the wrong
bound-state parameters.

In recent publications \cite{lms_prl,lmss_lcsr,lms_jphysg} we formulated a new 
algorithm for extracting the parameters of the ground state.
The idea proposed in these papers is to relax the standard assumption of a Borel-parameter-independent 
effective continuum threshold and allow for a Borel-parameter-dependent quantity.

In the present paper we develop this idea and provide details of
its application to the extraction of hadron parameters from
dispersive sum rules. We show that (i) the new algorithm leads to
improvements --- and for the case of the form factors to 
{\it significant} improvements --- of the actual accuracy of the
extracted bound-state parameters from all kinds of correlators in
the potential model; (ii) the {\it procedures} of extracting the
ground-state parameters from the truncated operator product
expansion (OPE) for the two-point function in potential models and
in QCD exhibit not only qualitative but also {\it quantitative}
similarities.

The paper is organized as follows: Section \ref{Sect:QM} presents
the application of our new algorithm to the extraction of the
ground-state parameters from correlators in a quantum-mechanical
potential model. Quantum-mechanical analogues of all correlators
used in QCD for the extraction of ground-state parameters (decay
constants and form factors) are discussed. Section \ref{Sect:QCD}
compares the extraction of the ground-state decay constant in QCD
and in quantum-mechanical potential models. Section \ref{Sect:Conclusions} 
presents our concluding remarks.

%\newpage
\section{\label{Sect:QM}Harmonic-oscillator potential model}
We consider a nonrelativistic Hamiltonian $H$ with a HO
interaction potential $V(r)$, $r\equiv|{\bf r}|$:
\begin{eqnarray}
H=H_0+V(r), \qquad H_0={\bf p}^2/2m, \qquad V(r)={m\omega^2r^2}/{2}.
\end{eqnarray}
In this model, all characteristics of the bound states are
easily calculable. For instance, for the ground state (g), one finds
\begin{eqnarray}
\label{EG}
E_{\rm g}=\frac{3}{2}\omega,\qquad R_{\rm g}\equiv |\psi_{\rm g}({\bf r=0})|^2=
\left(\frac{m\omega}{\pi}\right)^{3/2},\qquad F_{\rm g}(q)=\exp(-q^2/4m\omega),
\end{eqnarray}
where the elastic form factor of the ground state is defined in terms of the ground-state wave
function $\psi_{\rm g}$ as
\begin{eqnarray}
\label{FG} F_{\rm g}(q)=\langle \psi_{\rm g}|J({\bf q})|\psi_{\rm
g}\rangle= \int d^3k\,\psi^\dagger_{\rm g}({\bf k})\psi_{\rm
g}({\bf k - q})= \int d^3r\, |\psi_{\rm g}({\bf r} )|^2e^{i{\bf q
r}}, \qquad q\equiv |{\bf q}|,
\end{eqnarray}
and the current operator $J({\bf q})$ is given by the kernel
\begin{eqnarray}
\label{J} \langle {\bf r'}|J({\bf q})|{\bf r}\rangle=
\exp(i{\bf q r})\delta^{(3)}({\bf r-r'}).
\end{eqnarray}

%\newpage
\subsection{\label{Sect:Pi}Polarization operator}
The quantum-mechanical analogue of the Borelized polarization operator \cite{svz} has the form
\begin{eqnarray}
\label{pi} \Pi(T)=\langle {{\bf r}_f=0}|\exp(- H T)|{{\bf
r}_i=0}\rangle.
\end{eqnarray}
For the HO potential, the analytic expression for $\Pi(T)$ is well-known \cite{nsvz}:
\begin{eqnarray}
\label{piexact}
\Pi(T)=\left(\frac{\omega m}{\pi}\right)^{3/2}
\frac1{\left[2\sinh(\omega T)\right]^{3/2}}.
\end{eqnarray}
Apart from the overall factor, $\Pi(T)$ is a function of one parameter $\omega T$.

Let us define the average energy of the polarization function as
\begin{eqnarray}
\label{energypi}
E_\Pi(T)\equiv -\partial_T\log \Pi(T)=\frac32 \omega \coth(\omega T),\qquad
\partial_T\equiv \frac{\partial}{\partial T}.
\end{eqnarray}
At $T=0$ both $\Pi(T)$ and $E_\Pi(T)$ diverge.

Making use of the ``hadronic'' bound states of the model, we obtain the following representation for $\Pi(T)$:
\begin{eqnarray}
\Pi_{\rm hadr}(T)=\Pi_{\rm g}(T)+\Pi_{\rm excited}(T),\quad
\Pi_{\rm g}(T)=R_{\rm g}\exp\left({-E_{\rm g} T}\right).
\end{eqnarray}
For large values of $T$ the contributions of the excited states to the correlator vanish and therefore
$E_\Pi(T)\to E_{\rm g}$ for $T\to\infty$. The deviation of the energy from $E_{\rm g}$ at finite values of $T$
measures the ``contamination'' of the correlator by the excited states.

The OPE series is the expansion of $\Pi(T)$ at small Euclidean time $T$:
\begin{eqnarray}
\label{piope}
\Pi_{\rm OPE}(T)=\left(\frac{m}{2\pi T}\right)^{3/2}
\left(1-\frac{1}{4}\omega^2T^2+\frac{19}{480}{\omega^4 T^4}
+\cdots \right).
\end{eqnarray}
For $\Pi(T)$ this expansion is equivalent to the expansion in powers of the interaction $\omega$.
The first term in (\ref{piope}), $\Pi_0(T)$, does not depend on the interaction and describes the free
propagation of the constituents. The rest of the series represents power corrections:
%, which may be obtained just as the difference between the exact correlator and the free-propagation term:
\begin{eqnarray}
\label{pipower} \Pi_{\rm power}(T)=\Pi_{\rm OPE}(T)-\Pi_{0}(T).
\end{eqnarray}
$\Pi_0(T)$ may be written as the spectral integral
\cite{lms_2ptsr}
\begin{eqnarray}
\label{pi0} \Pi_0(T)=\int_0^\infty dz e^{-z T} \rho_0(z), \qquad
\rho_0(z)=\frac{m^{3/2}}{\sqrt{2}\pi^2}\sqrt{z},
\end{eqnarray}
with $\rho_0(z)$ the spectral density of the one-loop two-point Feynman diagram of the nonrelativistic
field theory \cite{lms_2ptsr}.

%\newpage
\subsection{\label{Sect:Gamma}Vertex function}
The basic quantity for the extraction of the form factor
in the method of dispersive sum rules is the correlator of three currents \cite{ioffe}.
The analogue of this quantity in quantum mechanics reads \cite{lms_3ptsr,radyushkin2001}
\begin{eqnarray}
\Gamma(E_2,E_1,q)= \langle {{\bf r}_f=0}|(H-E_2)^{-1}J({\bf
q})(H-E_1)^{-1}| {{\bf r}_i=0}\rangle,\qquad q\equiv |{\bf q}|,
\end{eqnarray}
[with the operator $J({\bf q})$ defined in (\ref{J})] and its
double Borel (Laplace) transform under $E_1\to \tau_1$ and $E_2\to\tau_2$
\begin{eqnarray}
\Gamma(\tau_2,\tau_1,q)= \langle {{\bf r}_f=0}|G(\tau_2)J({\bf q}
)G(\tau_1)|{{\bf r}_i=0}\rangle, \qquad G(\tau)\equiv \exp(-H
\tau).
\end{eqnarray}
For large $\tau_1$ and $\tau_2$ the correlator is dominated by the ground state:
\begin{eqnarray}
\Gamma(\tau_2,\tau_1,q)\to |\psi_{\rm g}({{\bf r}=0})|^2
e^{-E_{\rm g}(\tau_1+\tau_2)}F_{\rm g}(q^2)+\cdots.
\end{eqnarray}
Let us notice the Ward identity which relates the vertex function at zero momentum to the polarization operator:
\begin{eqnarray}
\label{pigamma}
\Gamma(\tau_2,\tau_1,q=0)=\Pi(\tau_1+\tau_2).
\end{eqnarray}
This expression follows directly from the current-conservation
relation
\begin{eqnarray}
J({{\bf q}=0})=1.
\end{eqnarray}
Let us consider the vertex function for equal times
$\tau_1=\tau_2=\frac12 T$, which in the HO model explicitly reads
\begin{eqnarray}
\label{gamma}
\Gamma(T,q)=\Pi(T)\exp\left(-\frac{q^2}{4m\omega}
\tanh\left(\frac{\omega T}{2}\right)\right).
\end{eqnarray}
The correlator $\Gamma$ is a function of two dimensionless variables $\omega T$ and $\hat q^2$,
$\hat q\equiv q/\sqrt{m\omega}$.

\begin{figure}
\begin{tabular}{cc}
\includegraphics[width=6cm]{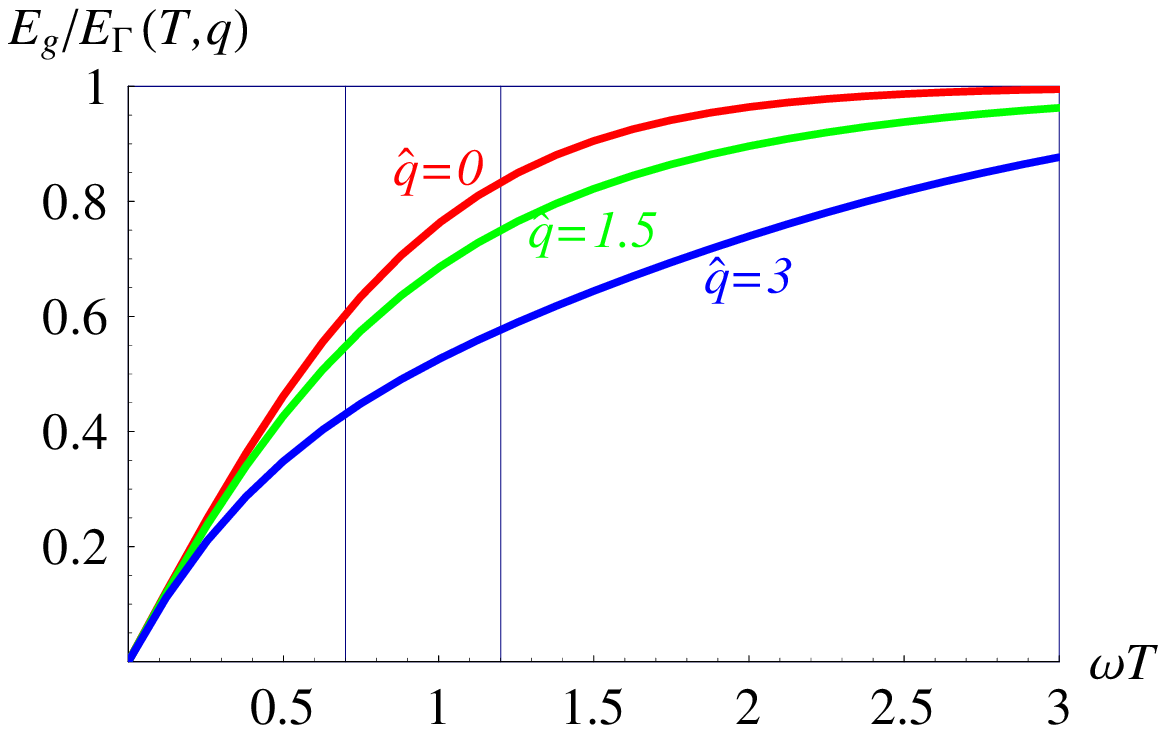}&
\includegraphics[width=6cm]{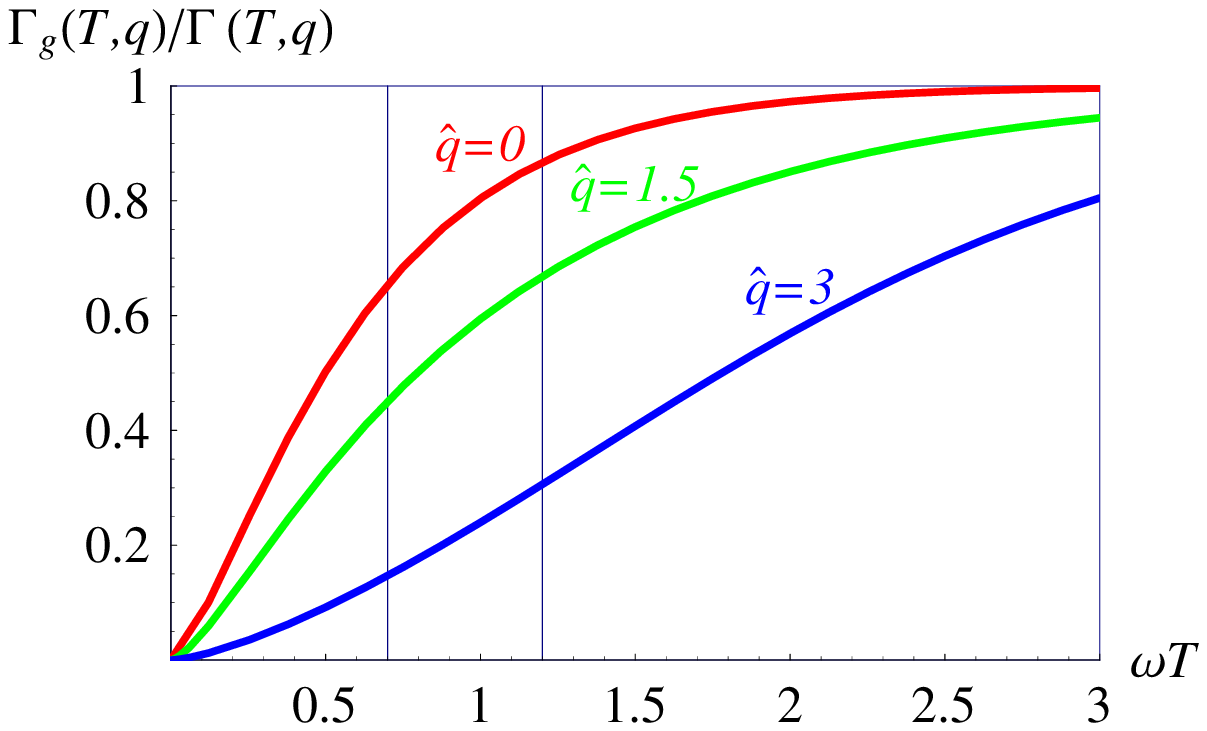}
\\
(a) & (b)
\end{tabular}
\caption{\label{Plot:1} Relative ground-state contribution to (a)
the energy $E_\Gamma(T,q)$ and (b) the correlator $\Gamma(T,q)$
vs. $T$ for several values~of~$q$. The vertical lines indicate the
boundaries of the ``Borel window''.}
\end{figure}

The corresponding average energy is defined as follows:
\begin{eqnarray}
\label{energygamma}
E_\Gamma(T,q)\equiv -\partial_T\log \Gamma(T,q)=\frac32 \omega \coth(\omega T)
+\frac{q^2}{4m}\frac{1}{\left(1+\cosh(\omega T)\right)}.
\end{eqnarray}
The contribution of the ground state to the correlator, $\Gamma_{\rm g}$, in the HO model has the form
\begin{eqnarray}
\label{gammaground}
\Gamma_{\rm g}(T,q)=R_{\rm g}\exp\left({-E_{\rm g} T}\right)F_{\rm g}(q^2).
\end{eqnarray}
The relative contribution of the ground state to $E_\Gamma(T,q)$
and $\Gamma(T,q)$ are shown in Fig.~\ref{Plot:1}: independently of
the value of $q$, $E(T,q)\to E_{\rm g}$ and $\Gamma(T,q)\to
\Gamma_{\rm g}(T,q)$ for large $T$. Notice, however, that the
saturation of the correlator by the ground state is delayed to
later $T$ with the growth of $q$.

Let us now construct for $\Gamma(T,q)$ the analogue of the OPE
as used in the method of three-point sum rules in QCD.
First, we expand $\Gamma$ in powers of $\omega^2$ and obtain
\begin{eqnarray}
\label{gamma1}
\Gamma(T,q)=\sum\limits_{n=0}^\infty\Gamma_{2n}(T,q)\omega^{2n}.
\end{eqnarray}
This expansion is the analogue of the expansion of the vertex
function in terms of the nonlocal condensates in QCD
\cite{nonlocal}. Explicitly, for the lowest terms one has
\begin{eqnarray}
\label{gamma11} \Gamma_0(T,q)&=&\left(\frac{m}{2\pi
T}\right)^{3/2}\exp\left(-\frac{q^2T}{8m}\right),\nonumber\\
\Gamma_2(T,q)&=&\left(\frac{m}{2\pi
T}\right)^{3/2}\exp\left(-\frac{q^2T}{8m}\right)
\left(-\frac14+\frac{q^2T}{96m}\right)\omega^2T^2,\quad \ldots.
\end{eqnarray}
The first term, $\Gamma_0$, corresponds to free propagation and does not depend on the confining potential.
It may be written as the double spectral representation \cite{lms_prd75}
\begin{eqnarray}
\Gamma_0(T,q)&=&\int_0^\infty dz_1 dz_2 e^{-{\frac12} z_1 T}
e^{-{\frac12} z_2 T}\Delta_0(z_1,z_2,q), \\ \nonumber
\Delta_0(z_1,z_2,q)&=&\frac{m^2}{4\pi^2
q}\theta\left(\left(z_1+z_2-\frac{q^2}{2m}\right)^2-4z_1z_2<0\right).
\end{eqnarray}
%with $\Delta_0$ the double spectral density of the triangle diagram $\Gamma_0$ in Fig.~\ref{Fig:1}.
The Ward identity in nonrelativistic field theory relates to each other the three-point function at
zero momentum transfer and the two-point function and leads to
\begin{eqnarray}
\label{wi}
\lim_{q\to 0}\Delta_0(z,z',q)=\delta(z-z')\rho_0(z).
\end{eqnarray}
Power corrections are defined as the difference between the exact expression and the free diagram:
\begin{eqnarray}
\label{gammaope}
\Gamma_{\rm OPE}(T,q)=\Gamma_0(T,q)+\Gamma_{\rm power}(T,q).
\end{eqnarray}
In order to obtain the analogue of the OPE for $\Gamma$ in terms
of local condensates, one should further expand $\Gamma_{\rm
power}(T,q)$ (i.e., $\Gamma_{2n}(T,q),$ $n\ge 1$) in powers of
$T$.

%\newpage
\subsection{\label{Sect:A}Vacuum-to-hadron correlator}
In order to apply the method of sum rules to hadron form factors
at intermediate and large momentum transfers a Borelized
vacuum-to-hadron amplitude of the $T$-product of two quark
currents is used \cite{stern}. The analogue of this quantity in
quantum mechanics has the form \cite{m_lcsr}
\begin{eqnarray}
\label{A} A(T,q)=\langle {{\bf r}=0}|G(T)J({\bf q})|\psi_{\rm
g}\rangle,\qquad G(T)\equiv \exp(-HT).
\end{eqnarray}
The spectral representation for $A(T,q)$ may be written in the form
\begin{eqnarray}
\label{disp}
A(T,q)=\int d^3k \;G(k^2,T)\psi_{\rm g}({\bf k}-{\bf q})=
\int_0^\infty dz\,\exp(-z T) \rho_A(z,T,q),\quad z\equiv {k}^2/2m,
\end{eqnarray}
where \cite{nsvz}
\begin{eqnarray}
\label{G} G(k^2,T)\equiv \langle {{\bf r}=0}|G(T)|{\bf k}\rangle=
\frac{1}{(2\pi)^{3/2}}\frac{1}{[\cosh(\omega
T)]^{3/2}}\exp\left(-\frac{k^2}{2 m \omega}\tanh(\omega T)\right).
\end{eqnarray}
Explicit expressions for $\rho_A(z,T,q)$ and $A(T,q)$ may be
obtained using $\psi_{\rm g}$ of Eq.~(\ref{EG}); for instance, for
$A(T,q)$~one~finds
\begin{eqnarray}
\label{Aexact}
A(T,q)=\sqrt{R_{\rm g}}\exp\left({-E_{\rm g} T}\right)
%\left(\frac{m\omega}{\pi}\right)^{3/4}\exp\left(-\frac{3}{2}\omega T\right)
\exp\left(-\frac{q^2}{4m\omega}\left(1-e^{-2\omega T}\right)\right).
\end{eqnarray}
The function $A(T,q)$ depends on two dimensionless variables 
${\hat q}^2= {q^2}/{m\omega}$ and $\omega T$.

Expanding in Eq.~(\ref{disp}) the Green function $G(k^2,T)$ in powers of $\omega$ generates the analogue
of the QCD twist expansion for the amplitude $A(T,q)$ (see \cite{m_lcsr} for details).

The ground-state contribution to the correlator has the form
\begin{eqnarray}
A_{\rm g}(T,q)=\sqrt{R_{\rm g}}\exp\left({-E_{\rm g} T}\right)F_{\rm g}(q),
\end{eqnarray}
so one finds
\begin{eqnarray}
\label{Rground}
\frac{A_{\rm g}(T,q)}{A(T,q)}=\exp\left(-\frac{q^2}{4m\omega}e^{-2\omega T}\right).
\end{eqnarray}
Notice the following features of the correlator $A(T,q)$ (see Fig.~\ref{Plot:2}):
\begin{figure}[!b]
\begin{tabular}{cc}
\includegraphics[width=6cm]{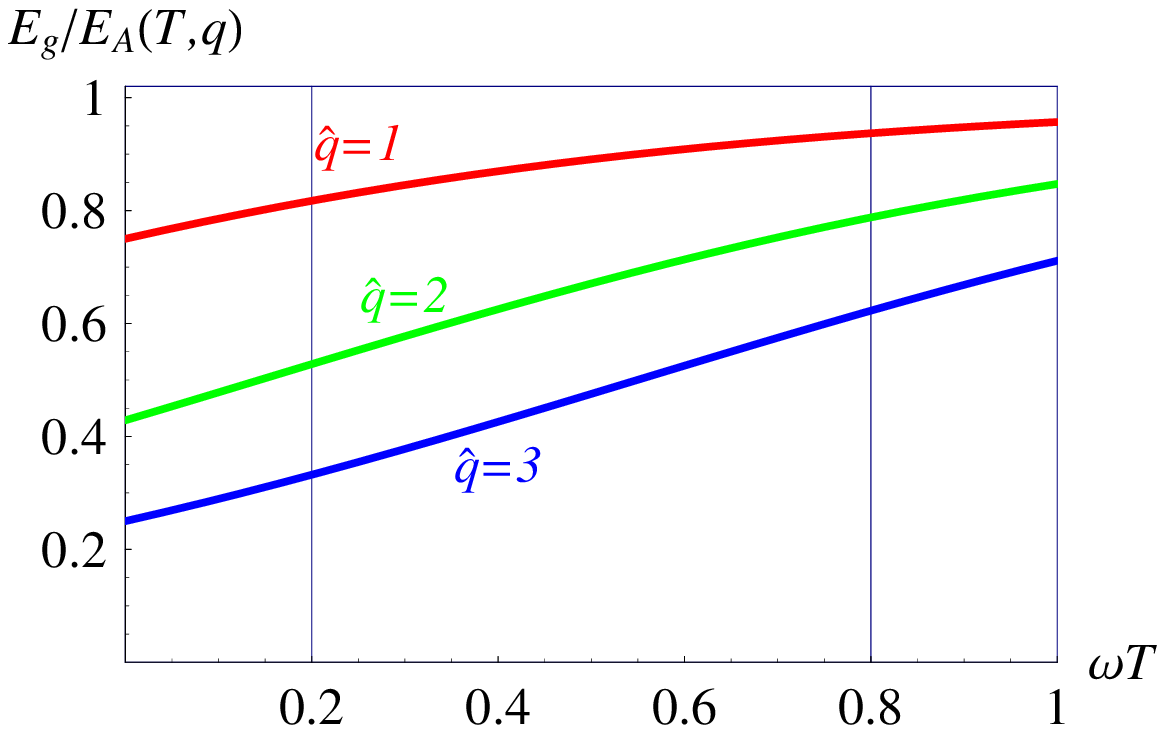}&
\includegraphics[width=6cm]{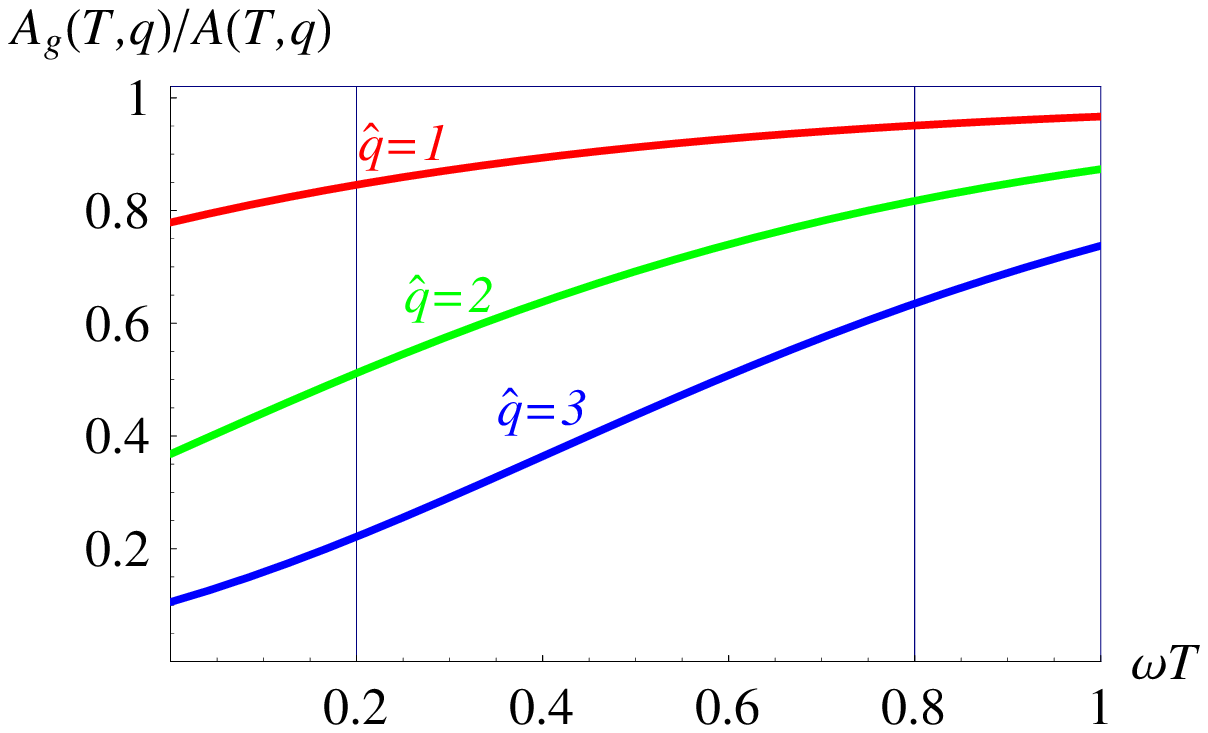}
\\
(a) & (b)
\end{tabular}
\caption{\label{Plot:2} Relative ground-state contribution to (a)
the energy $E_A(T,q)$ and (b) the correlator $A(T,q)$ vs. $T$ for
several values~of~$q$. The vertical lines indicate the boundaries
of the ``Borel window''. }
\end{figure}

\noindent
(i) At large $T$, the ground state provides the dominant contribution to $A$, similar to the case of
vacuum-to-vacuum correlators:
\begin{eqnarray}
A(T,q)\to A_{\rm g}(T,q) \quad \mbox{for} \quad T\to \infty.
\end{eqnarray}
(ii)
At small $q$, because of current conservation, the ground state dominates the correlator for all $T$.
This is a specific feature of $A$ which arises due to the choice of the initial state.
Thus, compared with vacuum-to-vacuum correlators, the correlator $A$ ``maximizes'' the ground-state contribution.

%%%%%%%%%%%%%%%%%%%%%%%%%%%%%%%%%%%%%%%%%%%%%%%%%%%%%%%%%%%%%%%%%%%%%%%%%%%%%%%%%%%

\subsection{Dual correlators and effective continuum thresholds}
The sum rule for each of the correlators described above is an
expression of the fact that the correlator calculated in terms of
the constituent degrees of freedom (quarks and gluons) and the
bound-state degrees of freedom (hadrons) are equal to each other.
In order to isolate the ground-state contributions from the sums
over hadron spectra, one invokes quark-hadron duality which
assumes that the excited states are dual to the high-energy
regions --- i.e., the regions above some {\it effective continuum
thresholds} --- of the spectral representations for the
correlators. The correlators to which the corresponding cuts have been 
applied are referred to as the {\it dual} correlators. The
relations between the ground-state contributions and the dual
correlators take the following form:
\begin{eqnarray}
\label{dual_pi}
R_{\rm g}e^{-{E_{\rm g}}T}&=&\Pi_{\rm dual}(T,z^\Pi_{\rm eff}(T))
%\nonumber\\&&
\equiv
\int\limits_{0}^{z^\Pi_{\rm eff}(T)}dz\,e^{-z T}\rho_0(z)+ \Pi_{\rm power}(T),
\\
\label{dual_gamma}
F_{\rm g}(q)R_{\rm g}e^{-{E_{\rm g}}T}&=&\Gamma_{\rm dual}(T,q,z^\Gamma_{\rm eff}(T,q))
%\nonumber\\&&
\equiv
\int\limits_{0}^{z^\Gamma_{\rm eff}(T,q)}dz_1 \int\limits_{0}^{z^\Gamma_{\rm eff}(T,q)}dz_2\,e^{-\frac12 (z_1+z_2)T}
\Delta_0(z_1,z_2,q)+ \Gamma_{\rm power}(T,q), %\nonumber\\
\\
\label{dual_a} F_{\rm g}(q)\sqrt{R_{\rm g}}e^{-{E_{\rm g}}T}&=&
A_{\rm dual}(T,q,z^A_{\rm eff}(T,q))
%\nonumber\\&&
\equiv\int_{0}^{z^A_{\rm eff}(T,q)}dz\,e^{- z T} \rho_A(z,T,q).
\end{eqnarray}
As consequence of (\ref{pigamma}), $\Pi_{\rm power}(T)=\Gamma_{\rm
power}(T,q=0)$ and
\begin{eqnarray}
z^\Pi_{\rm eff}(T)=z^\Gamma_{\rm eff}(T,q=0).
\end{eqnarray}
Then, due to the Ward identity (\ref{wi}), the relations (\ref{dual_pi}) and
(\ref{dual_gamma}) yield the correct normalization of the form factor
$F_{\rm g}(q=0)=1$, as required by current conservation.

\begin{figure}[!b]
\begin{tabular}{cc}
\includegraphics[width=6cm]{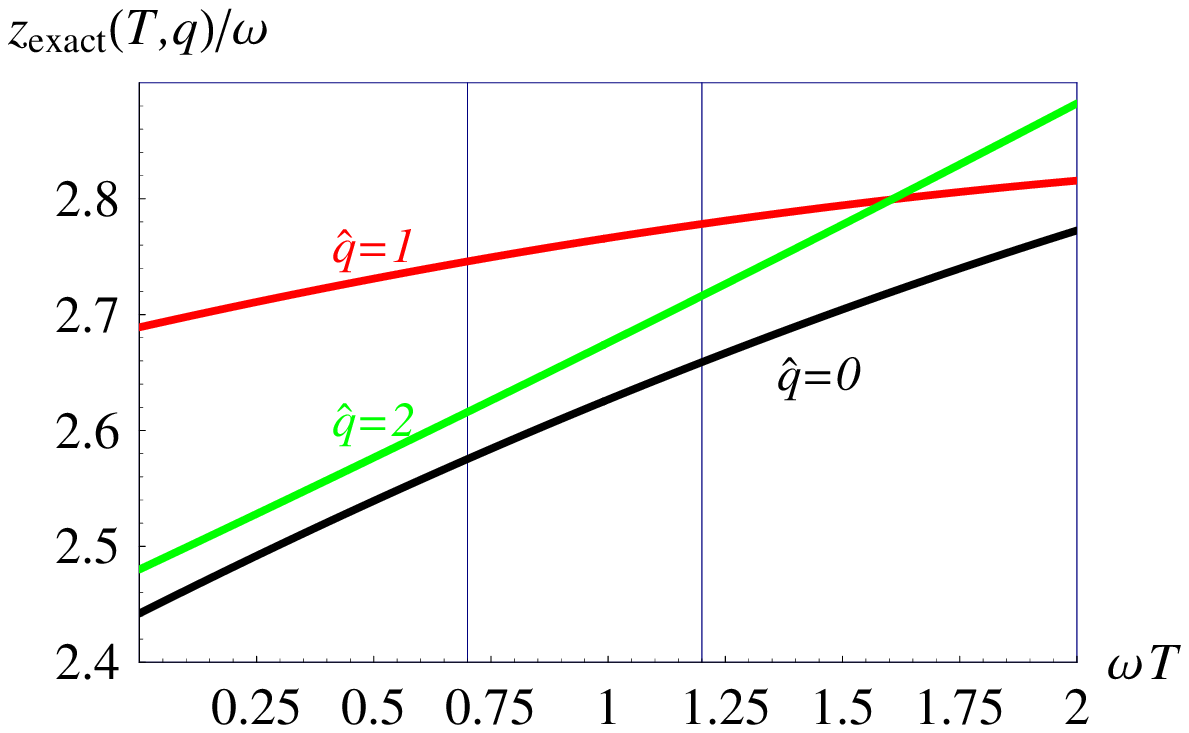}
&
\includegraphics[width=6cm]{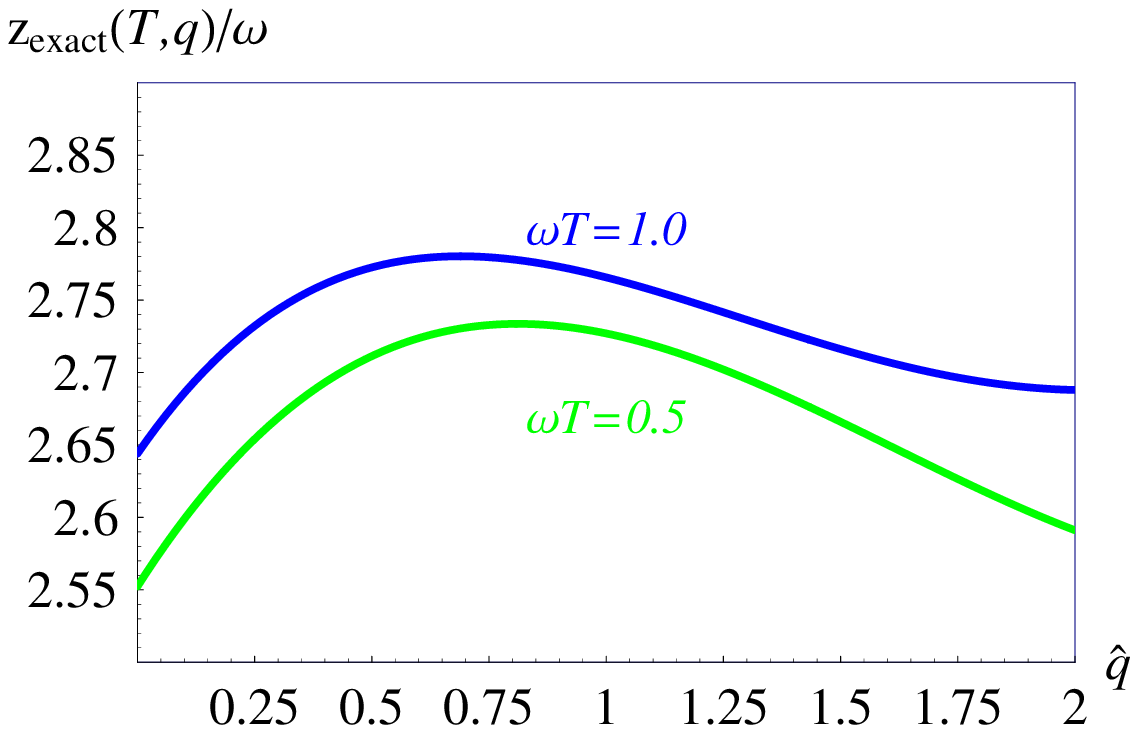}
\\
(a) & (b)
\\
\includegraphics[width=6cm]{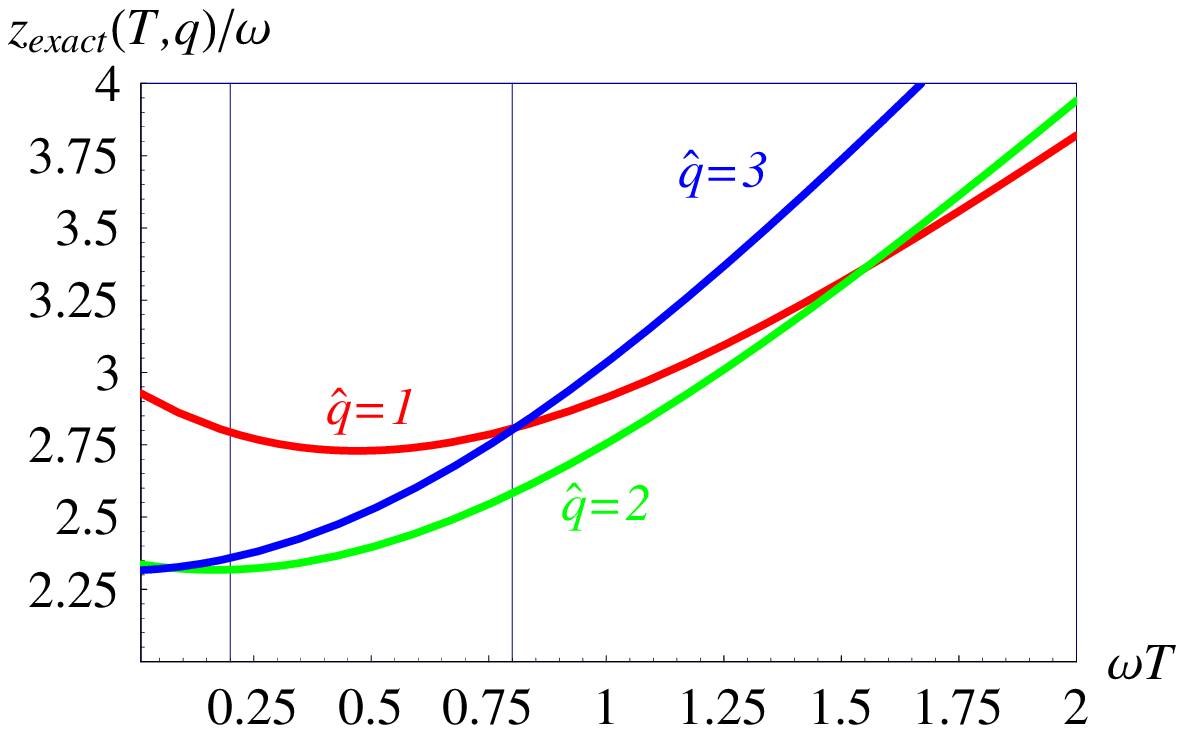}
&
\includegraphics[width=6cm]{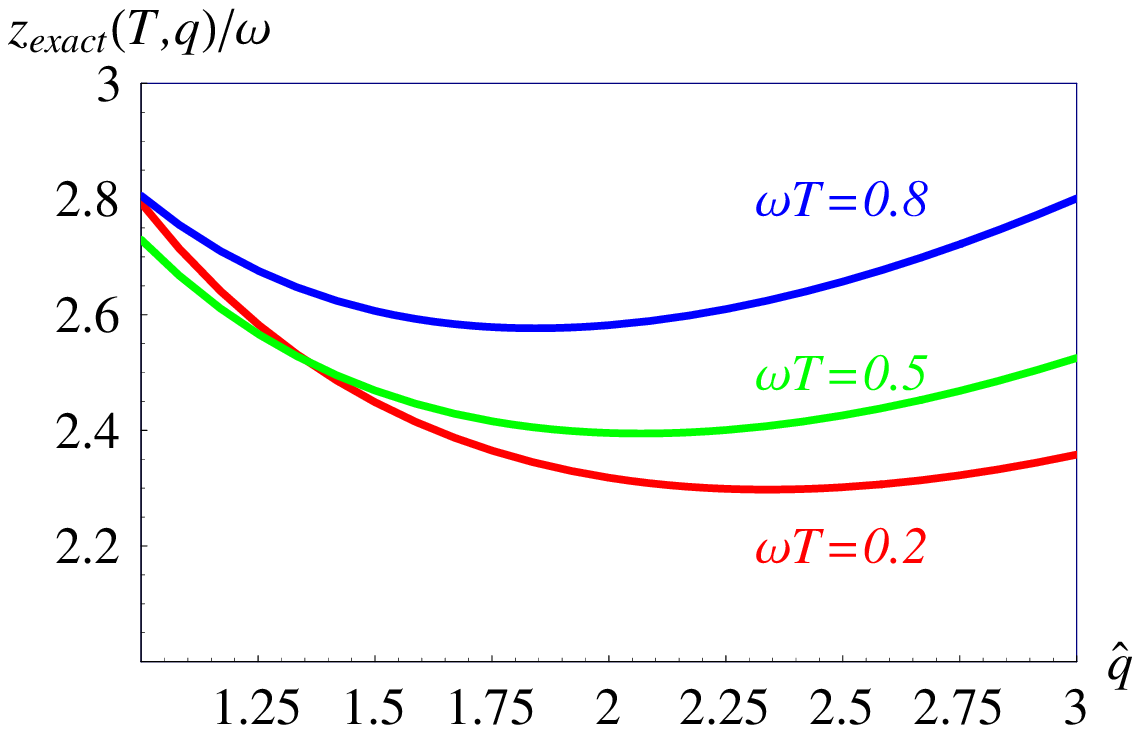}
\\
(c) & (d)
\\
\end{tabular}
\caption{\label{Plot:3} Exact effective continuum threshold
$z^\Gamma_{\rm exact}(T,q)$ (a,b) and $z^A_{\rm exact}(T,q)$
(c,d). $z^\Pi_{\rm exact}(T)=z^\Gamma_{\rm exact}(T,q=0)$. The
vertical lines indicate the ``Borel window''.}
\end{figure}

Let us take advantage of knowing the exact ground-state parameters
in the potential model and calculate the {\it exact effective
continuum thresholds} for all the correlators. These exact
thresholds are obtained by solving the equations above for 
the exact ground-state parameters on the l.h.s.; the exact thresholds make the
quark-hadron duality relations (\ref{dual_pi})--(\ref{dual_a})
exact. Clearly, $z^\Gamma_{\rm exact}(T,q=0)=z^\Pi_{\rm
exact}(T)$. Figure~\ref{Plot:3} depicts the exact effective
continuum thresholds for $\Gamma(T,q)$ and $A(T,q)$.
As is obvious from Fig.~\ref{Plot:3}, the effective continuum
thresholds do depend considerably on both $T$ and $q$. Moreover,
for the same theory (i.e., the same potential and the same values
of its parameters) the exact effective thresholds for
$\Gamma(T,q)$ and $A(T,q)$ are rather different from each other
and have very different $T$- and $q$-dependencies.

It might be useful to notice that the dependence of the effective
thresholds on the Borel parameter, being unusual for spectral
representations, does not contradict any properties of field
theory: the dual correlators are hand-made objects and they cannot
be written as, e.g., the $T$-product of local currents; thus the
dual correlators have completely different analytic properties
than the ``usual'' correlators in field theory.\footnote{The dual
correlator has a relatively simple form in the Borel-parameter
space; the dual correlator in the momentum space is a nonlocal
object. This is not strange: the dual correlator should reproduce
a single ground-state pole by the dispersion integral over a
finite energy region. Therefore, the dependence of the effective
threshold on $T$ is its inherent property required by the analytic
properties of the dual correlator.}

To summarize, the exact effective continuum thresholds (i) depend
on $T$ and on $q$; (ii) are not ``universal'', i.e., for one and
the same ground state, the effective thresholds in different
correlators strongly differ from each other. As a consequence,
making use of the effective continuum threshold obtained, e.g.,
for $\Pi_{\rm dual}$ as effective thresholds for the dual
correlators $\Gamma_{\rm dual}$ or $A_{\rm dual}$, as often done
in practical applications of the sum-rule method in QCD, has no
ground and may lead to very inaccurate estimates for the form
factors.

On the practical side, a good or a bad extraction of the ground-state parameters depends on our capacity to find
a reasonable approximation to the exact effective continuum threshold.

%%%%%%%%%%%%%%%%%%%%%%%%%%%%%%%%%%%%%%%%%%%%%%%%%%%%%%%%%%%%%%%%%%%%%%%%%%%%%%%%%%%
%\newpage

\subsection{Numerical analysis}
Let us consider a restricted problem where the energy $E_{\rm g}$ of the ground state and its wave
function at the origin, $\sqrt{R_{\rm g}}$, are known, and try to extract its elastic form factor
from the sum rules (\ref{dual_gamma}) and (\ref{dual_a}).

A. First, according to \cite{svz} we should determine the working
interval of $T$ --- the ``Borel window'' --- where the method may
be applied to the extraction of the ground-state parameter:

\noindent
(i) The upper boundary of the $T$-window is obtained from the condition that the
truncated expansion gives a good approximation to the exact correlator.
Since in the HO model the correlators are known exactly, the upper boundary
is $T = \infty$.
However, to be close to realistic situations, where only a limited number of higher-order or higher-twist corrections is available,
following our results reported in \cite{lms_3ptsr,m_lcsr} we define the upper boundary of the window as
$\omega T \lesssim 1.2$ for $\Gamma(T,q)$ and $\omega T \lesssim 0.8$ for $A(T,q)$.

\noindent (ii) The lower boundary of this $T$-window is determined
by the condition that the ground state gives a ``sizable''
contribution to the correlator. If we require the ground-state
contribution to exceed, say, 50\%, the window for $\Gamma(T,q)$
closes already for $\hat q \simeq 2$ (Fig.~\ref{Plot:1}b), whereas
for $A(T,q)$ it closes for $\hat q \simeq 4$ (Fig.~\ref{Plot:2}b).

We shall see, however, that our algorithm will allow us to extract the form factor,
although with a worse accuracy, also in the region where the ground-state contribution to the correlator
remains well below 50\%. Thus, our algorithm opens the possibility to study also the region of larger momentum transfers.

We choose the ``windows'' as follows:
\begin{eqnarray}
&&0.7 \lesssim \omega T \lesssim 1.2 \quad \mbox{ for }
\Gamma(T,q),
\\ &&0.2 \lesssim \omega T \lesssim 0.8 \quad \mbox{ for } A(T,q).
\end{eqnarray}

B. Second, we must choose a criterion to fix the effective continuum threshold $z_{\rm eff}(T, q)$.
We proceed in the following way: We consider a set of $T$-dependent Ans\"atze for the effective 
continuum threshold
\begin{eqnarray}
\label{zeff}
z^{(n)}_{\rm eff}(T, q)= \sum\limits_{j=0}^{n}z_j^{(n)}(q)(\omega T)^{j}.
\end{eqnarray}
(The standard procedure adopted in all sum-rule applications in QCD is to assume
a $T$-independent quantity. Obviously, this option is also included in our analysis.)
Now, at each value of $q$ we fix the parameters on the
r.h.s of (\ref{zeff}) as follows: we calculate the dual energy
\begin{eqnarray}
\label{Edual} E^\Gamma_{\rm dual}(T, q) &=& - \frac{d}{d T} \log
\Gamma_{\rm dual}(T, q, z^\Gamma_{\rm eff}(T, q)),
\\
E^A_{\rm dual}(T, q) &=& - \frac{d}{d T} \log A_{\rm dual}(T, q,
z^A_{\rm eff}(T, q)),
\end{eqnarray}
for the $T$-dependent $z_{\rm eff}$ of Eq.~(\ref{zeff}). Then we
evaluate $E_{\rm dual}(T, q)$ at several values of $T = T_i$ ($i =
1,\dots, N$, where $N$ can be taken arbitrarily large) chosen
uniformly in the window. Finally, we minimize the squared
difference between $E_{\rm dual}$ and the exact value $E_{\rm g}$:
\begin{eqnarray}
\label{chisq}
\chi^2 \equiv \frac{1}{N} \sum_{i = 1}^{N} \left[ E^{(A,\Gamma)}_{\rm dual}(T_i, q) - E_{\rm g}\right]^2.
\end{eqnarray}
Recall that the exact effective continuum thresholds for the two
correlators $\Gamma(T,q)$ and $A(T,q)$ have rather different
$T$-dependencies, see Fig.~\ref{Plot:3}. Therefore, one should
expect that the expressions (\ref{zeff}) for the corresponding
effective thresholds will be different.

\begin{figure}
\begin{tabular}{cc}
\includegraphics[width=6cm]{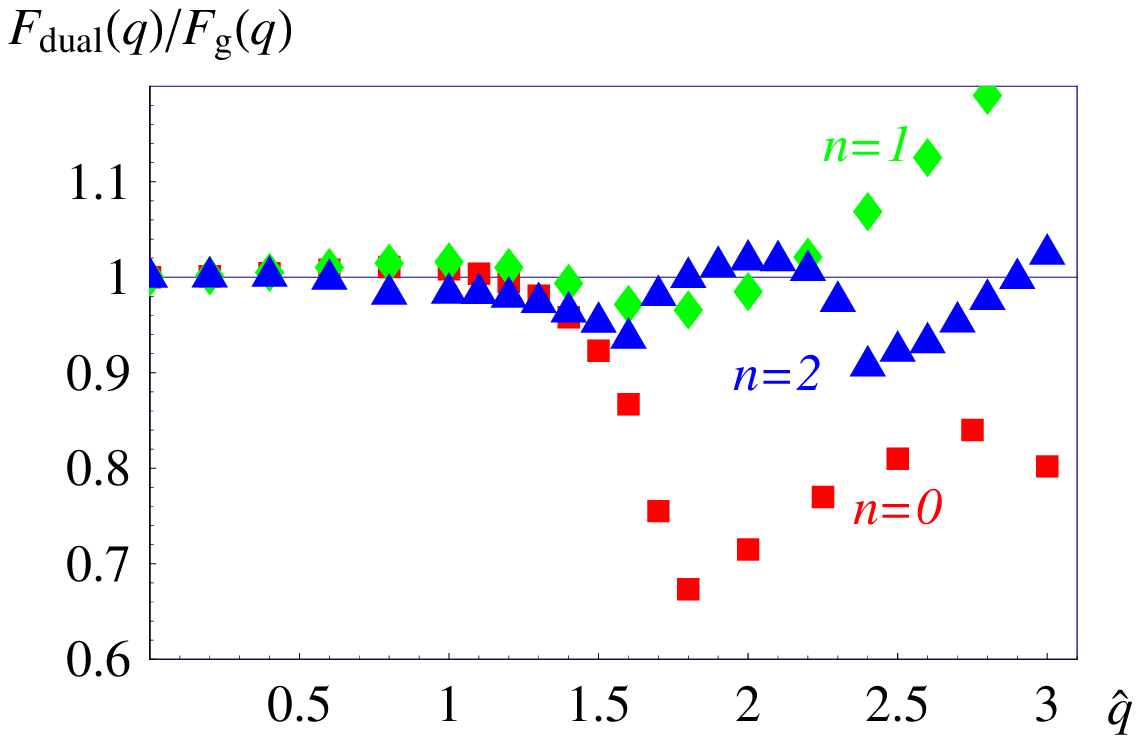}&
\includegraphics[width=6cm]{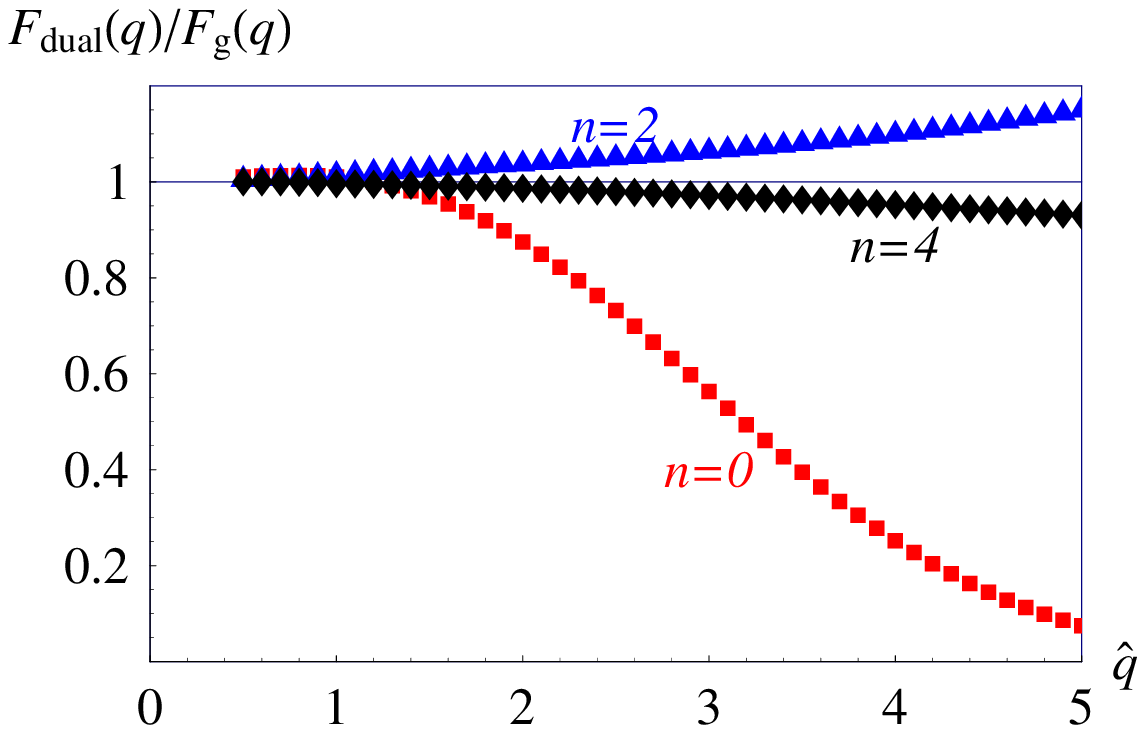}\\
(a) & (b)
\end{tabular}
\caption{\label{Plot:4} The dual form factor $F_{\rm dual}(q)$
extracted from (a) the correlator $\Gamma(T,q)$ and (b) the
correlator $A(T,q)$. Red symbols ($n=0$) correspond to the
standard $T$-independent Ansatz for $z_{\rm eff}$, green ($n=1$)
to the linear Ansatz, blue ($n=2$) to the quadratic Ansatz, and
black ($n=4$) to the quartic Ansatz.}
\end{figure}

The results for the form factor obtained from $\Gamma(T,q)$ and
from $A(T,q)$ upon optimizing the parameters of $z^\Gamma_{\rm
eff}$~and~$z^A_{\rm eff}$ according to (\ref{chisq}) are shown in
Fig.~\ref{Plot:4}.

\subsubsection{Form factor from $\Gamma(T,q)$}
Let us first consider the extraction of the form factor from the correlator $\Gamma(T,q)$.

Within the standard $T$-independent approximation, one extracts
the form factor in the region $\hat q\le 1.5$ with better than
10\% accuracy, while the accuracy decreases rather fast for higher
$q$ (see Fig.~\ref{Plot:4}a). The dramatic increase of the error
at $\hat q \gtrsim 2$ is related to the fact that the contribution
of the ground state to the correlator decreases rapidly with $q$
in the given $T$-window $0.7\le \omega T\le 1.2$ (see
Fig.~\ref{Plot:1}).

\begin{figure}[!b]
\begin{tabular}{cc}
\includegraphics[width=6cm]{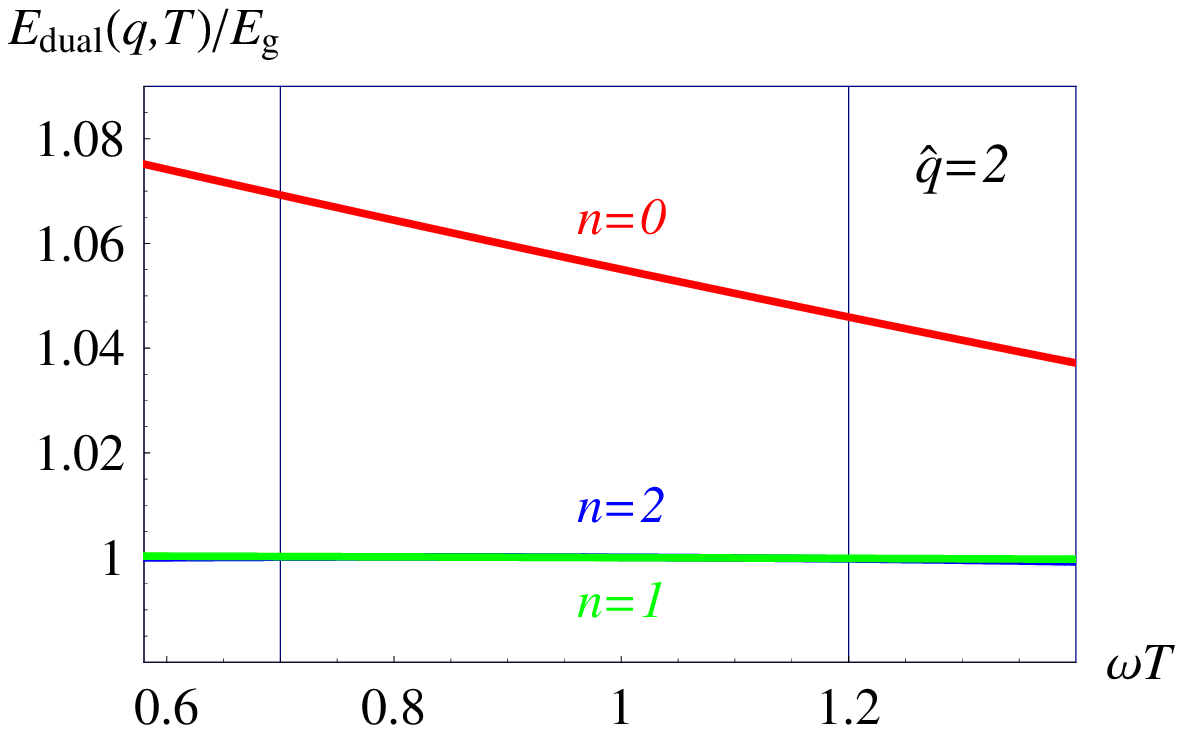}&
\includegraphics[width=6cm]{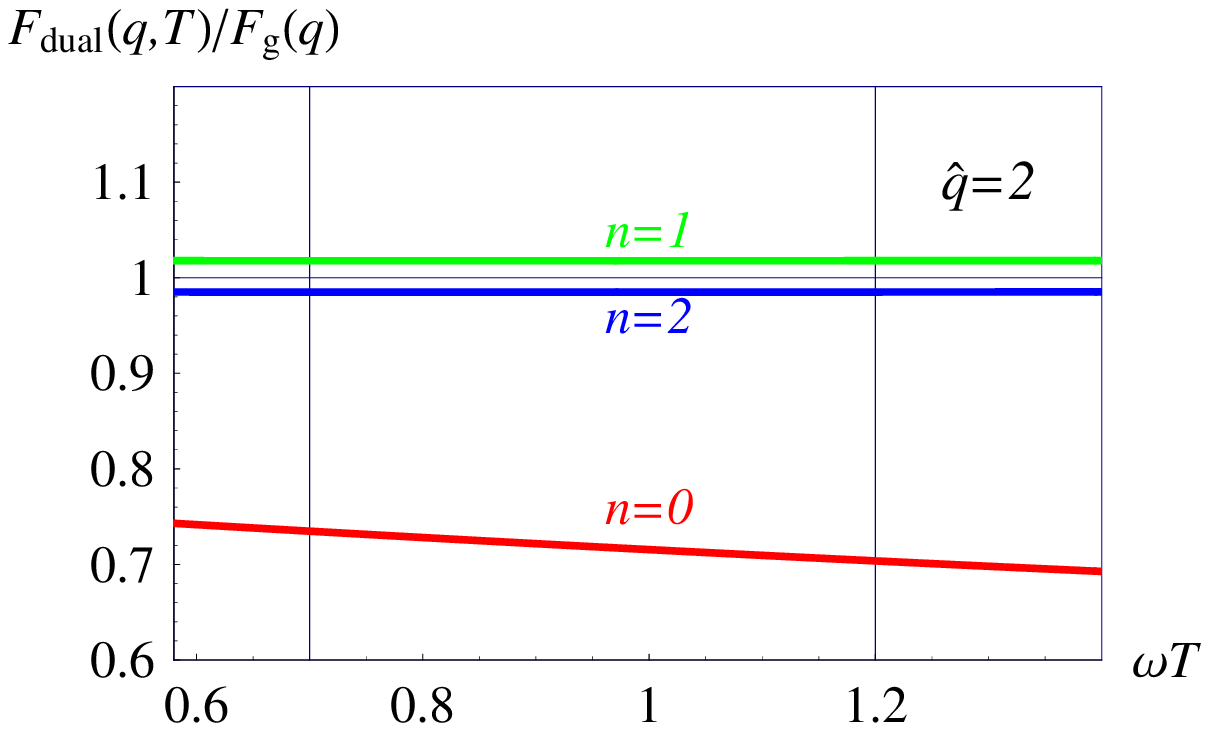}
\\
(a) & (b)
\end{tabular}
\caption{\label{Plot:5} The correlator $\Gamma(T,q)$: the fitted
dual energy $E_{\rm dual}(T,q)$ and the corresponding dual form
factor $F_{\rm dual}(T,q)$ for $\hat q=2$. Red lines ($n=0$)
correspond to the standard $T$-independent Ansatz for $z_{\rm
eff}$, green lines ($n=1$) to the linear Ansatz, and blue lines
($n=2$) to the quadratic Ansatz.}
\end{figure}

The real problem is, however, that the magnitude of the error cannot be obtained
on the basis of the standard criteria adopted in the method: e.g., at $\hat q=2$, the variation of the extracted form
factor in the window is only about 2\% (see Fig.~\ref{Plot:5}b) which mimics an accurate extraction of the form factor, whereas
the actual error comprises 25\%.
Thus, within a $T$-independent Ansatz for the effective threshold, the error of the form factor cannot be determined
from the variation of the dual form factor in the window.

%%%%%%%%%%%%%%%%%%%%%%%%%%%%%%%%%%%%%%%%%%%%%%%%%%%%%%%%%%%%%%%%%%%%%%%%%%
%%% GAMMA
Allowing for the $T$-dependent effective threshold leads to a
considerable improvement of the accuracy: both for the linear and
the quadratic Ans\"atze for the threshold, the dual energy
reproduces the known energy rather accurately in the window, and
the dual form factor reproduces the exact form factor rather well.
Noteworthy, the {\it actual} error of the extracted form factor
may be obtained as the spread between the dual form factors
obtained for the linear and the quadratic Ans\"atze for $z_{\rm eff}$.

\subsubsection{Form factor from $A(T,q)$}
Let us now consider the extraction of the form factor from the correlator $A(T,q)$.

\begin{figure}[!ht]
\begin{tabular}{cc}
\includegraphics[width=6cm]{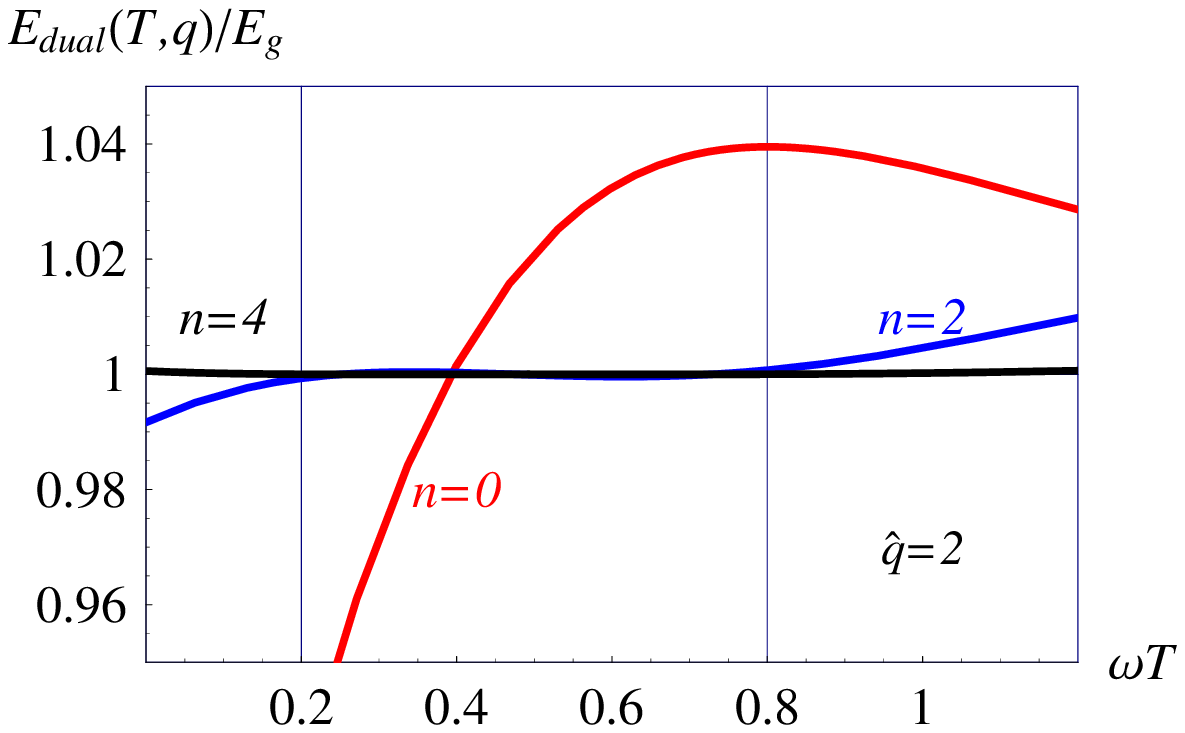}&
\includegraphics[width=6cm]{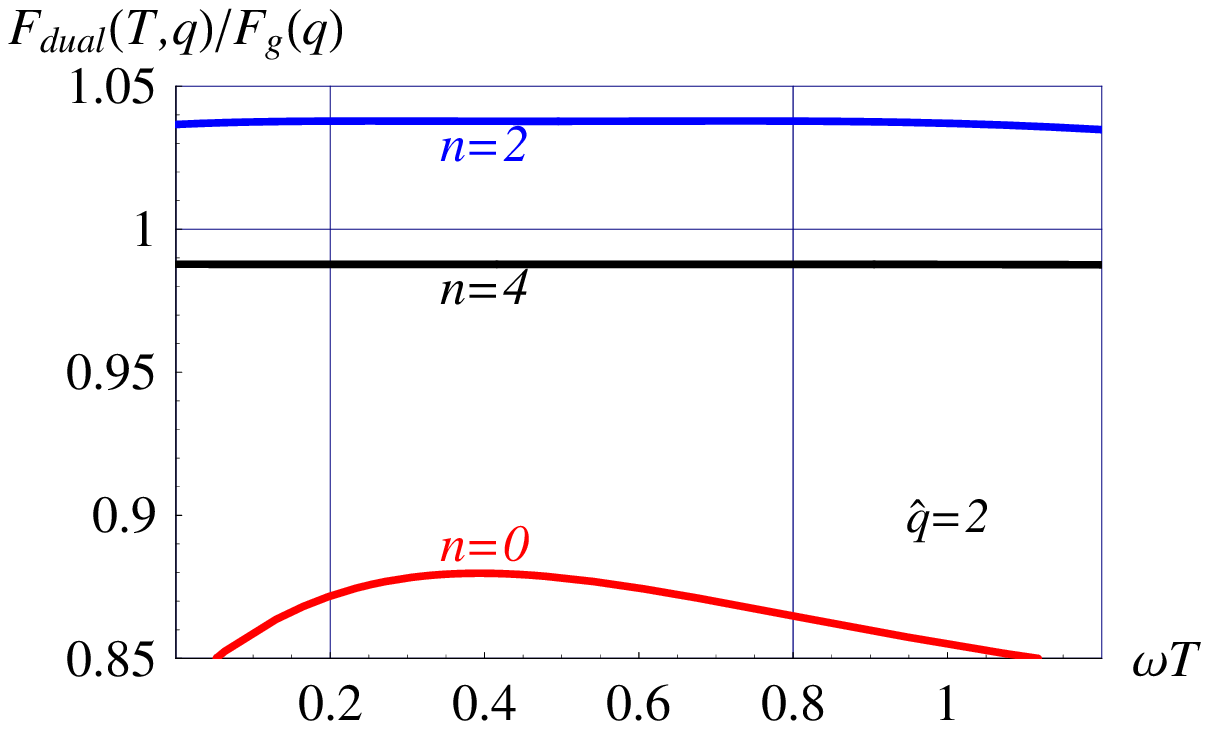}
\end{tabular}
\caption{\label{Plot:6} The correlator $A(T,q)$: the fitted dual
energy $E_{\rm dual}(T,q)$ and the corresponding dual form factor
$F_{\rm dual}(T,q)$ for $\hat q=2$. Red lines ($n=0$) correspond
to the standard $T$-independent Ansatz for $z_{\rm eff}$, blue
lines ($n=2$) to the quadratic Ansatz; and black lines ($n=4$) to
the quartic Ansatz. }
\end{figure}

The standard $T$-independent Ansatz works well below $\hat q\le 1.5-2$, but for 
larger $\hat q$ the accuracy of the extracted form
factor decreases very fast, see Fig.~\ref{Plot:4}. The main
problem here is the same as for the case of $\Gamma(T,q)$: the
large error in the extracted form factor cannot be guessed on the
basis of the standard Borel stability criterion: as seen in
Fig.~\ref{Plot:6}, the variation of the form factor at $\hat q=2$
in the window comprises only 2\%, whereas the actual error exceeds
10\%.

Allowing for $T$-dependent approximations for $z_{\rm eff}$ leads to considerable improvements: as shown in
Fig.~\ref{Plot:4},
the form factor may be extracted with a reasonable accuracy (better than 15\%) up to $\hat q=5$
with the quadratic and quartic Ans\"atze, $z^{(2)}_{\rm eff}(T, q)$ and $z^{(4)}_{\rm eff}(T, q)$
(the results obtained for $n=1$ are close to the results for $n=0$, and the results for $n=3$ are close to those for $n=2$).

This high accuracy is achieved in spite of the fact that (i) the
relative contribution of the ground state to the correlator falls
down considerably below 50\% in the $T$-window, and (ii) the form
factor at $\hat q=5$ falls down compared to its value at $\hat
q=0$ by almost three orders of magnitude.

Notice again the typical feature of the extraction procedure: both
for $n=2$ and $n=4$ both the dual energy $E_{\rm dual}(T,q)$ and
the dual form factor $F_{\rm dual}(T,q)$ are extremely stable
(better than $0.1 \%$) in the window for all $\hat q\le 5$
(Fig.~\ref{Plot:6} shows the results for $\hat q=2$). In spite of
this stability, the actual error of the extracted form factor for
$n=2$ and $n=4$, separately, is found to be much larger: e.g., at
$\hat q=5$ this error comprises 10--15\%. Again, this error could
not be guessed on the basis of the standard Borel stability
criterion, which thus does not work also for vacuum-to-hadron
correlators.

Nevertheless, we have seen clear improvements in the results of
the extraction procedure as soon as one goes beyond the assumption
of a $T$-independent effective continuum threshold: First, the
actual accuracy turns out to be (much) better. Second, the
$T$-dependent Ansatz allows one to extract the form factor in a
broader range of the momentum transfer. Finally, the spread
between the results obtained with the quadratic and the quartic
Ans\"atze for $z_{\rm eff}$ provides the {\it actual} error of the
extracted form factor.

%%%%%%%%%%%%%%%%%%%%%%%%%%%%%%%%%%%%%%%%%%%%%%%%%%%%%%%%%%%%%%%%%%%%%%%%%%
%%%%% A

%%%%%%%%%%%%%%%%%%%%%%%%%%%%%%%%%%%%%%%%%%%%%%%%%%%%%%%%%%%%%%%%%%%%%%%%%%%
%\newpage
\section{\label{Sect:QCD}Potential model vs. QCD}
In the previous section we have shown that the concept of the $T$-dependent effective continuum threshold and the way
of fixing this quantity by reproducing the bound-state energy has led to a dramatic increase of the actual accuracy
of the extracted form factor in the potential model. 

Does this new approach to the extraction of the properties of
ground-state hadrons promise equivalent improvements in QCD? We
give a positive answer to this question \cite{lms_qmvsqcd}.
\begin{figure}[!ht]
\begin{tabular}{cc}
\includegraphics[width=6cm]{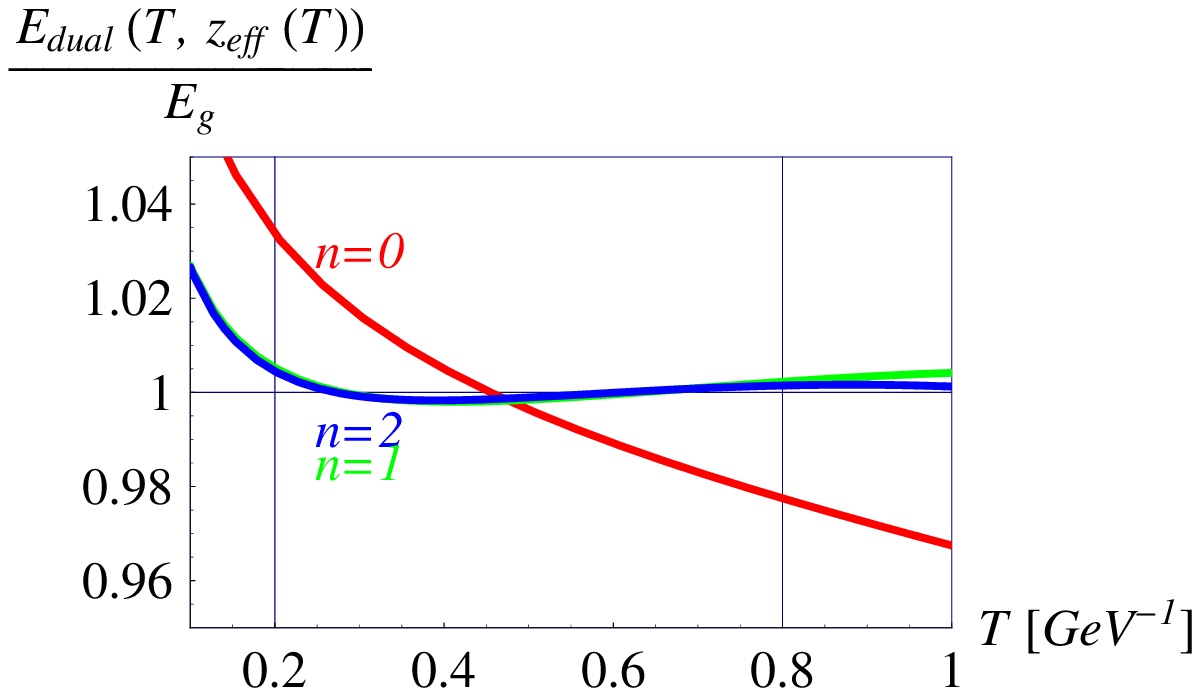}& \includegraphics[width=6cm]{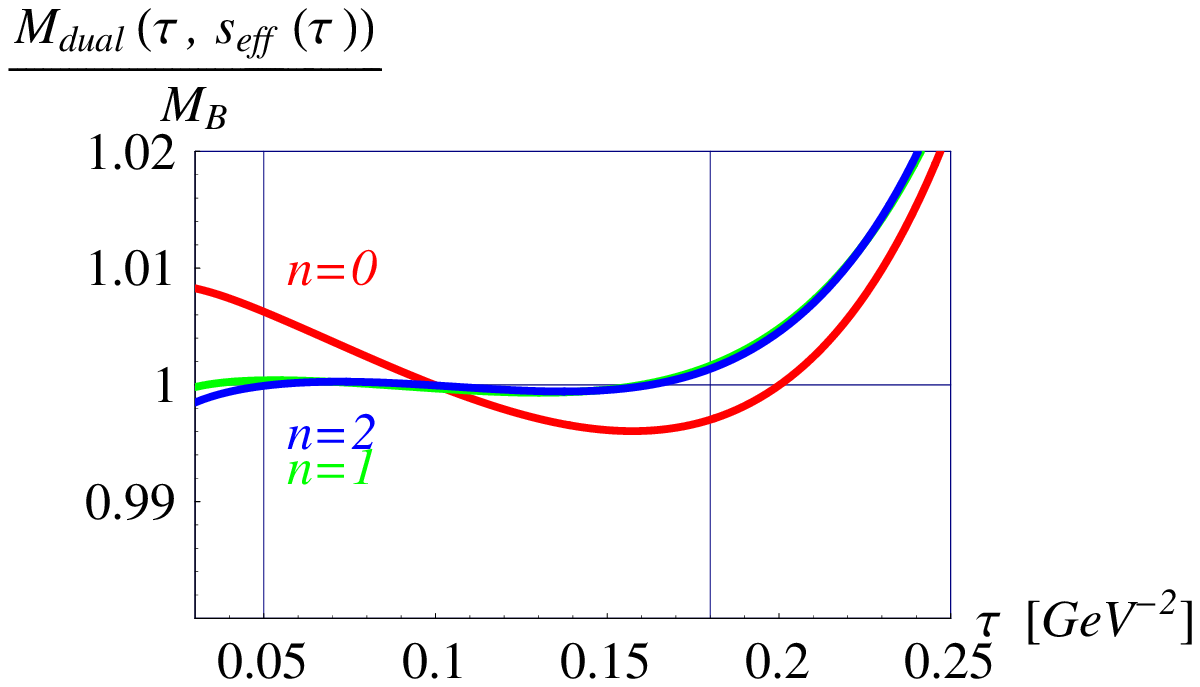}\\
\includegraphics[width=6cm]{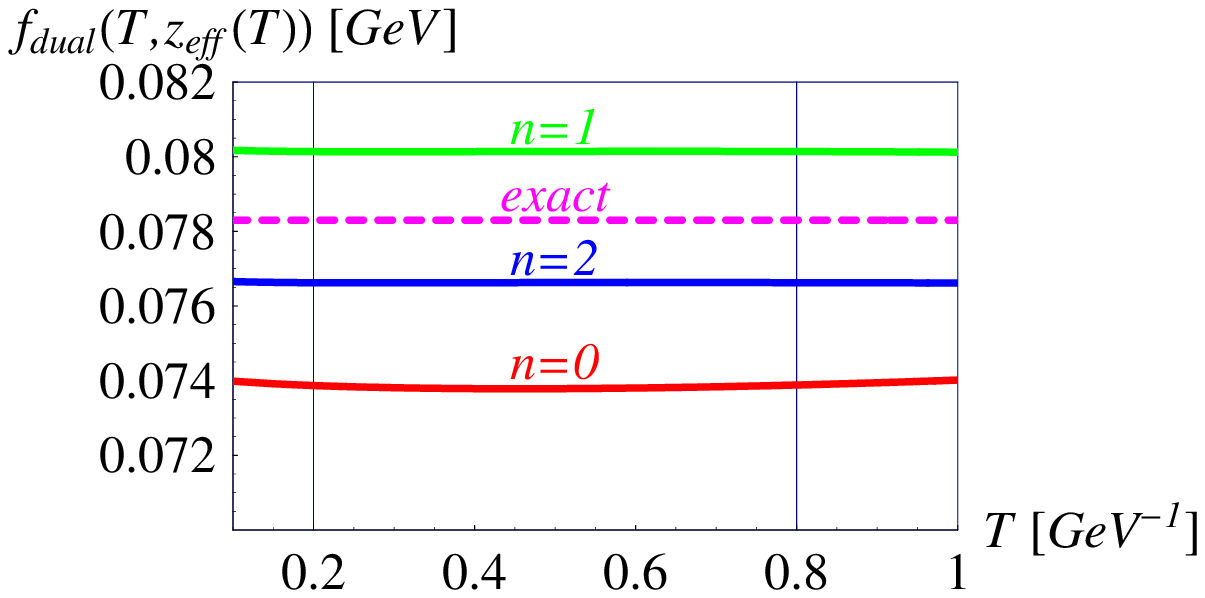}& \includegraphics[width=6cm]{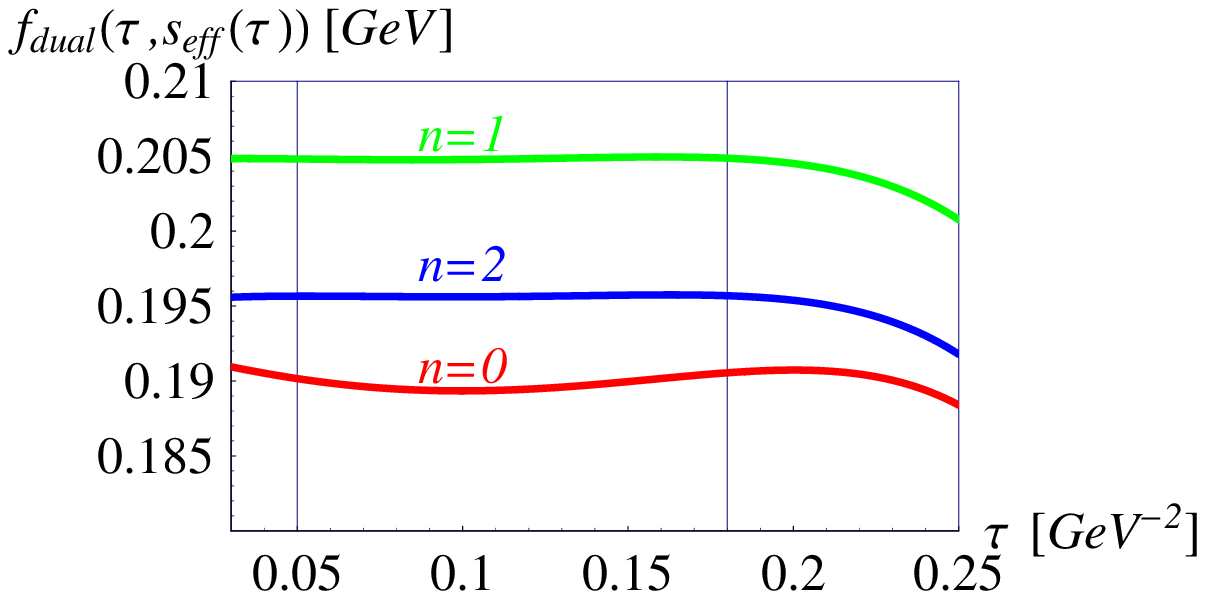}
\end{tabular}
\caption{\label{Plot:7} Dual mass and extracted dual decay
constant vs.\ Borel parameter in the potential model (left column)
and in QCD (right column). In QCD, the Borel parameter $\tau$ has dimension $GeV^{-2}$. 
}
\end{figure}
Figure~\ref{Plot:7} compares the extraction of the ground-state
decay constant from the two-point function in the potential model (left column) 
and in QCD (right column). 
For the potential model, we consider the interaction consisting of a confining HO part and a Coulomb
part and make use of the OPE which includes the radiative corrections up to
order $\alpha^2$ and several lowest-order power corrections. This
corresponds to the best known OPE for the two-point function of
heavy-light quark currents \cite{jamin} which we use for QCD. 
Details of this 
analysis including the specific values of the numerical parameters
can be found in \cite{lms_qmvsqcd}; they are not essential for our
purpose here. 

From Fig.~\ref{Plot:7} it becomes evident that --- with respect to
the {\it extraction} of the ground-state parameters --- there are
no essential differences between QCD and quantum mechanics: as
soon as the parameters of the Lagrangian are fixed, and the
truncated OPE is calculated with a reasonable accuracy (taking
into account also the relevant choice of the renormalization scale
in QCD), the extraction procedures are very similar.

At first glance, this similarity might look surprising since the
structure of bound states in quantum-mechanical potential models
and in QCD are rather different. However, the method of dispersive
sum rules does not make use of the details of the ground-state
structure. What really matters for extracting the ground-state
parameters by this method is (i) the way one implements the
quark-hadron duality and (ii) the structure of the OPE for a given
correlator. Therefore, it should not be surprising that the
extraction procedures in potential models and in QCD are similar,
as soon as quark-hadron duality is implemented in both theories in the
same way by a cut on the spectral representation for the
correlators.

In the potential model, the actual value of the decay constant
belongs to the band provided by the linear and the quadratic
Ans\"atze for the effective continuum threshold. It seems very
likely that the same happens in QCD, at least this may be taken as
a realistic conjecture. Perhaps even more important is that
studying the $\tau$-dependent parameterizations for the effective
threshold allows one to probe the magnitude of the intrinsic,
i.e., systematic, uncertainty of the extracted decay constant.
Interestingly, the systematic uncertainties of the ground-state
parameters are probed solely by the OPE, without making use of the
information about the excited states of the hadron spectrum. The
intrinsic uncertainty should be taken into account along with the
``usual'' uncertainty obtained in the method of QCD sum rules by
the variation of the input QCD parameters; while the latter is going
to decrease when increasing the accuracy of these parameters,
e.g., $m_b(m_b)$, the systematic uncertainty does not depend on
the input QCD parameters and remains at the same level.

%\newpage
\section{\label{Sect:Conclusions}Conclusions}
We studied the extraction of the ground-state form factor from the
vacuum-to-vacuum and vacuum-to-hadron correlators. Let us
summarize the main messages of our analysis:

\begin{itemize}
\item
%$\bullet$ 1.
%A.
The knowledge of the correlator in a limited range of relatively small Euclidean
times $T$ (i.e., large Borel masses) is not sufficient for the determination of the
ground-state parameters. In addition to the OPE for the relevant correlator, one needs an
independent criterion for fixing the effective continuum threshold.

\item
%$\bullet$ 2.
%B.
Assuming a $T$-independent (i.e., a Borel-parameter-independent)
effective continuum threshold, the error of the extracted dual
form factor $F_{\rm dual}(T,q)$ turns out to be typically much
larger than the variation of $F(T,q)$ in the Borel window.
Therefore the standard procedures based on the Borel stability
adopted in the method of sum rules do not allow one to obtain
realistic error estimates for the extracted form factor;
eventually the actual error of the extracted form factor may happen 
to be much larger.

\item
%$\bullet$ 3.
%C.
In those cases where the ground-state energy (mass) is known,
considerable improvements may be achieved by allowing for a
$T$-dependent effective continuum threshold and finding its
parameters by minimizing the deviation of the energy of the dual
correlator from the known value of the ground-state energy,
Eq.~(\ref{chisq}). First, the actual accuracy of the extracted
form factor is considerably improved. Second, the spread between
form-factor~values obtained with different $T$-dependent Ans\"atze
for $z_{\rm eff}$ provides a realistic error of the form factor.

\item
%$\bullet$ 3.
%D.
We have demonstrated the striking similarity --- not only
qualitatively but, more importantly, also quantitatively ---
between the extraction procedures in the quantum-mechanical
potential model and in QCD. This similarity is the consequence of
the same way of implementing the quark-hadron duality in both
theories. The established similarity strongly suggests that the
results on the extraction of ground-state form factors obtained in
potential models have direct implications for the corresponding
analyses in QCD and should be taken very seriously.

Although a rigorous --- in the mathematical sense --- control over
the systematic errors of the extracted hadron parameters remains
unfeasible, the proposed modifications related to the
Borel-parameter-dependent effective continuum threshold promise a
considerable increase of the {\it actual} accuracy of the method
of sum rules and opens the possibility to obtain {\it realistic\/}
estimates for the systematic errors.

\end{itemize}

\vspace{.5cm}
{\it Acknowledgments.}
We are grateful to Hagop Sazdjian for the most pleasant collaboration on several issues
discussed in this paper.
D.~M. gratefully acknowledges financial support from the Austrian Science Fund (FWF) under project P20573
and from the President of Russian Federation under grant for leading scientific schools 1456.2008.2.

\end{document}